\documentclass[fleqn,10pt]{wlscirep}
\usepackage{graphicx}
\usepackage{epsfig}
\usepackage{epstopdf}
\usepackage{amsmath}

\title{Orbital-dependent charge dynamics in MnP revealed by optical study}

\author[1,*,$\ddagger$]{P. Zheng}
\author[1,2,$\ddagger$]{Y. J. Xu}
\author[1,2,$\ddagger$]{W. Wu}
\author[1]{G. Xu}
\author[1,2]{J. L. Lv}
\author[1,2]{F. K. Lin}
\author[1]{P. Wang}
\author[1,2,3,$\dagger$]{Yi-feng Yang}
\author[1,2,3]{J. L. Luo}
\affil[1]{Beijing National Laboratory for Condensed Matter Physics and Institute of Physics, Chinese Academy of Sciences, Beijing 100190, China}
\affil[2]{School of Physical Sciences, University of Chinese Academy of Sciences, Beijing 100190, China}
\affil[3]{Collaborative Innovation Center of Quantum Matter, Beijing 100190, China}

\affil[*]{pzheng@iphy.ac.cn}
\affil[$\dagger$]{yifeng@iphy.ac.cn}
\affil[$\ddagger$]{These authors contributed equally to this work.}

\begin{abstract}
Unconventional superconductivity often emerges at the border of long-range magnetic orders. Understanding the low-energy charge dynamics may provide crucial information on the formation of superconductivity. Here we report the unpolarized/polarized optical conductivity study of high quality MnP single crystals at ambient pressure. Our data reveal two types of charge carriers with very different lifetimes. In combination with the first-principles calculations, we show that the short-lifetime carriers have flat Fermi sheets which become gapped in the helimagnetic phase, causing a dramatic change in the low-frequency optical spectra, while the long-lifetime carriers are anisotropic three-dimensional like which are little affected by the magnetic transitions and provide major contributions to the transport properties. This orbital-dependent charge dynamics originates from the special crystal structure of MnP and may have an influence on the unconventional superconductivity and its interplay with helimagnetism at high pressures.
\end{abstract}

\begin{document}

\flushbottom
\maketitle

\thispagestyle{empty}

\section*{Introduction}
Magnetic quantum fluctuations at the border of long-range antiferromagnetic or ferromagnetic orders have been generally believed to be the major pairing force for unconventional superconductivity. The discoveries of superconductivity in the helical magnets CrAs \cite{Wu, Kotegawa} and MnP \cite{Cheng} reveal a fascinating new family of superconductivity emerging at the border of long-range helimagnetic orders. As the first manganese-based superconductor, MnP has a helimagnetic ground state at ambient pressure. Gradually applying pressure on good single crystals first destroys the helimagnetic order at about $1.2\,$GPa and then yields a second helimagnetic phase at higher pressures as revealed very recently by synchrotron-based magnetic X-ray diffraction techniques, neutron diffraction, muon-spin rotation and nuclear magnetic resonance (NMR) \cite{Wang, Matsuda, Yano, Yano2, Khasanov, Fan}. Near the critical pressure of $7.8\,$GPa, where the second helimagnetic phase is also suppressed, superconductivity emerges with $T_c\approx1\,$K. Above $T_c$, the resistivity exhibits non-Fermi liquid behavior and reveals a dramatic enhancement of the quasiparticle effective mass at low temperatures in the normal state, suggesting an unconventional pairing mechanism for the observed superconductivity at the border of the helimagnetic phase \cite{Cheng}. Understanding the relation between superconductivity and helimagnetism is an important issue in this material.

To shed light on the nature of the charge dynamics and its interplay with magnetism, we performed infrared optical measurements on MnP at temperatures from 300 K to 10 K at ambient pressure, where a para-ferromagnetic transition has been reported at about $T_{C1}=289.5-292\,$K and a ferro-helimagnetic transition at $T_{C2}=47-53\,$K in previous studies \cite{Takase1, Huber, Felcher, Yamazaki, Cheng, Zieba, Todate, Shiomi, Reis}. In the helimagnetic state, the Mn spins rotate within the $ab$ plane with a propagation vector $Q_{h}$ along the $c$-axis \cite{Felcher}. The compound remains metallic at all temperatures and no structural change has been reported with temperature. The temperature-dependent optical investigation hence allows us to reveal the behavior of the charge carriers in response to the complicated magnetic orders. We find that the low-frequency metallic responses in the ferromagnetic phase can be best understood as a sum of two Drude components, one of which has a very short lifetime and becomes gapped along the $b$-axis in the helimagnetic phase, causing a dramatic change in the low-frequency optical spectra crossing the ferro-helimagnetic phase transition. This indicates two types of charge carriers with very different lifetimes. In combination with the first-principles calculations, we identify that the short-lifetime carriers have flat Fermi sheets showing quasi-one-dimensional (1D) character near the Fermi energy and contribute mostly to the magnetic properties, while the long-lifetime carriers have anisotropic three-dimensional (3D) Fermi surfaces and are largely responsible for the transport properties such as the $dc$ conductivity. Our results reveal an orbital-dependent helimagnetic phase transition and suggest a microscopic two-fluid picture for the charge dynamics in MnP. These put a constraint on any serious microscopic model and may have an influence on the interplay between unconventional superconductivity and helimagnetism at high pressures.

\section*{Results}

\begin{figure}[h]
\centering
\includegraphics[width=0.5\textwidth]{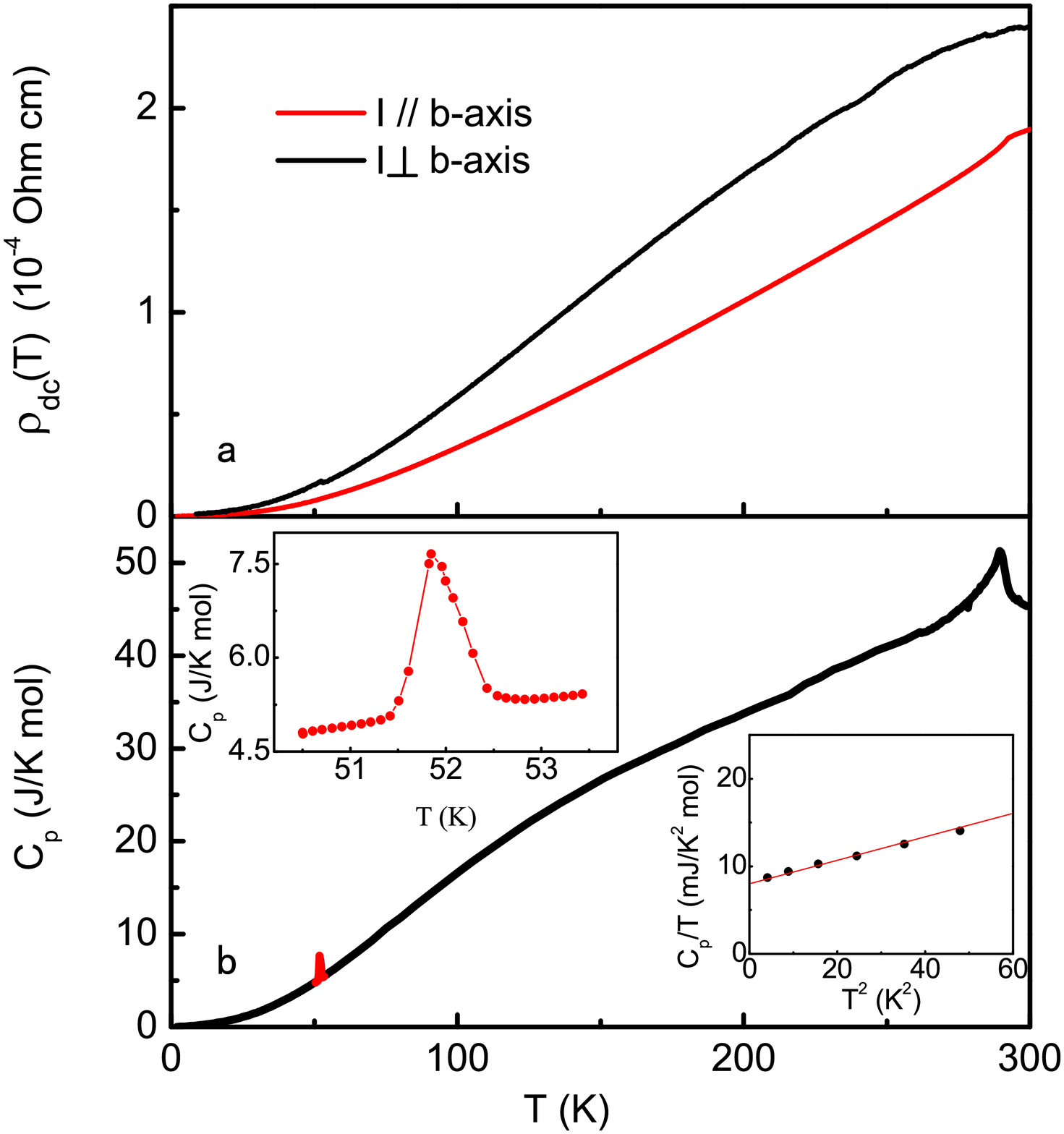}
\caption{\label{fig1}(Color online) \textbf{Resistivity and specific heat of MnP}. (a) Temperature dependence of the $dc$ resistivity $\rho_{dc}$(T) along and perpendicular to the $b$-axis; (b) Temperature dependence of the specific heat $C_p(T)$, showing a para-ferromagnetic transition at about $290\,$K and a ferro-helimagnetic transition at about $51.8\,$K. The insets show the specific heat peak at the ferro-helimagnetic phase transition in the enlarged scale and the fit of $C_p(T)/T$ versus $T^2$ at low temperatures.}
\end{figure}

\textbf{Transport and heat capacity measurements.} Figure \ref{fig1}a shows a typical result of the $dc$ resistivity $\rho_{dc}(T)$ along and perpendicular to the $b$-axis. We see that the compound remains metallic at all temperatures and our sample has a very high RRR ratio, $\rho_{dc}(300\,{\rm K})/\rho_{dc}(2\,{\rm K})\approx 1000$, similar to that of the superconducting sample \cite{Cheng}. The temperature dependence of the specific heat is plotted in Fig. \ref{fig1}b. A para-ferromagnetic transition and a ferro-helimagnetic transition are observed at about $290\,$K and $51.8\,$K, respectively. As shown in the inset, a tiny sharp peak is seen for the first time at the ferro-helimagnetic phase transition, owing to the high quality of our single crystals \cite{Stolen, Takase2}. A tiny signature is also observed in the resistivity data ($I\perp b$) in Fig. \ref{fig1}a. At low temperatures, the specific heat can be nicely fitted with $C_p/T=\gamma_{e}+\beta T^{2}$, where $\gamma_{e}\approx 8.3\,$mJ/mol K$^2$ is the residual electron specific heat coefficient and the $\beta T^{2}$ term is the phonon contribution.\\

\begin{figure}[h]
\centering
\includegraphics[width=0.5\textwidth]{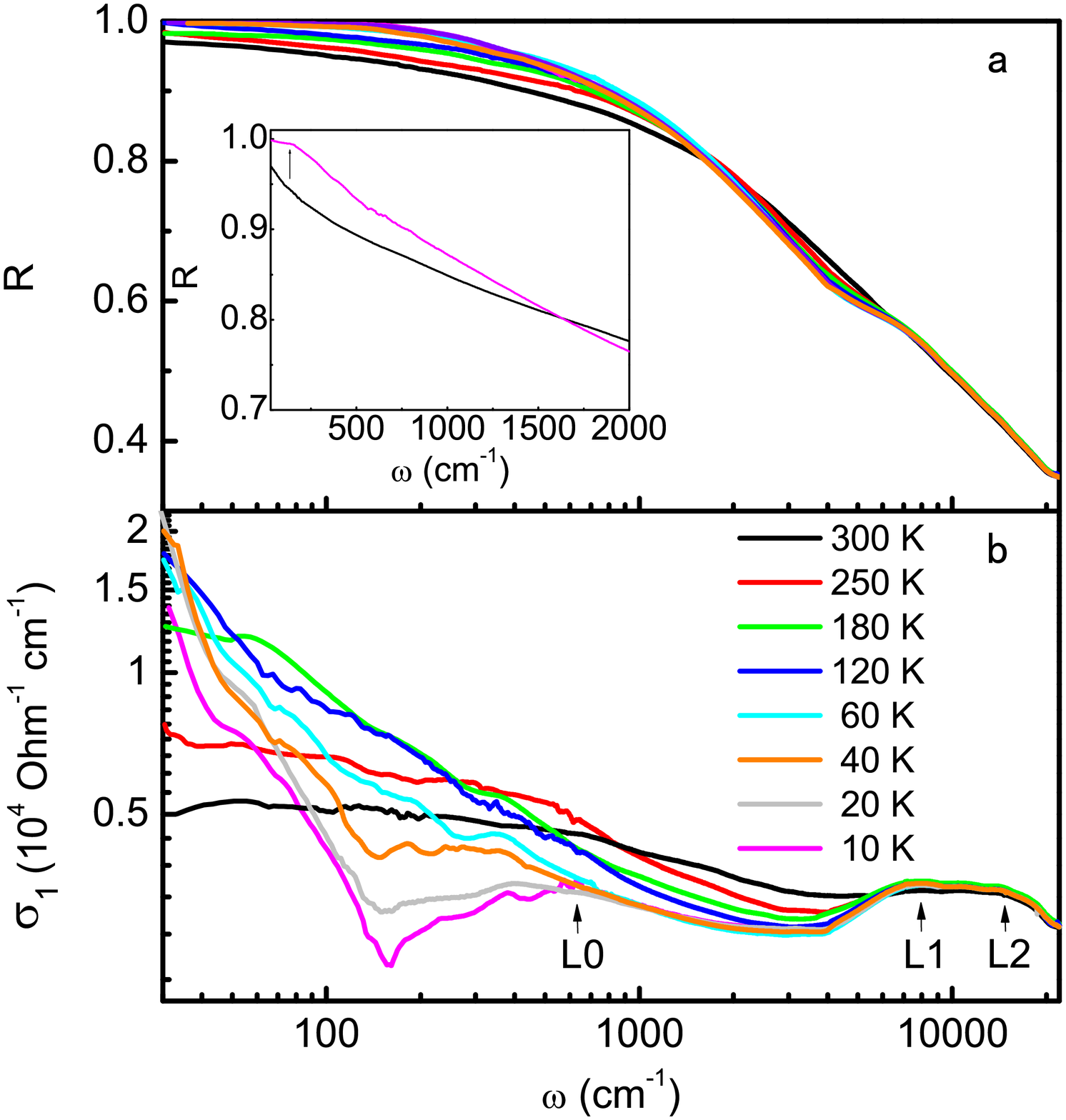}
\caption{\label{fig2}(Color online) \textbf{Unpolarized optical reflectivity and conductivity of MnP.} (a) Temperature dependence of the reflectivity spectra $R(\omega)$ measured with unpolarized light. The  arrow in the inset denotes the low-$\omega$ reflectance edge at $10\,$K. (b) Temperature dependence of the real part of the optical conductivity $\sigma_1(\omega)$. The two peaks at high frequencies remain almost unchanged with temperature. }
\end{figure}

\textbf{Unpolarized optical spectra.} Figure \ref{fig2} plots the optical reflectivity $R(\omega)$ measured from $30$ cm$^{-1}$ to $25000$ cm$^{-1}$ with unpolarized light. It is seen that as temperature decreases, $R(\omega)$ increases in the low frequency region, but decreases at higher frequency region below $8000\,$cm$^{-1}$. All the spectra intersect between $1000\,$cm$^{-1}$ and $2000\,$cm$^{-1}$. At low temperatures, a broad peak emerges at about $8000\,$cm$^{-1}$, above which all spectra merge together and decrease monotonically with increasing $\omega$, while the low-frequency reflectivity, as shown in the inset of Fig. \ref{fig2}a, increases faster towards unity than those at higher temperatures, giving rise to a low-$\omega$ reflectance edge at 10 K in the optical conductivity. The above features are more pronounced in the real part of the optical conductivity, $\sigma_1(\omega)$, as plotted in Fig. \ref{fig2}b for temperatures ranging from $10\,$K to $300\,$K. The optical conductivity $\sigma_1(\omega)$ was derived from the Kramers-Kronig (KK) transformation after using the Hagen-Rubens relation for the low-frequency extrapolation of $R(\omega)$. In the high frequency region, the reflectivity was extrapolated as a constant to $\omega=80 000\,$cm$^{-1}$, above which a well-known function of $\omega$$^{-4}$ was used.

Two broad peaks are seen at around $8100\,$cm$^{-1}$ and $16000\,$cm$^{-1}$ (denoted as L1 and L2), independent of temperature. These peaks may be attributed to electronic correlations or inter-band transitions between the bonding/antibonding bands of Mn-3$d$ and P-3$p$ orbitals \cite{Yanase, Bonfa, Grosvenor}. For intermediate frequencies between $1500\,$cm$^{-1}$ and $5000\,$cm$^{-1}$, the spectra depend weakly on frequency and decrease gradually with lowering temperature. In the low frequency region, on the other hand, a peak is seen to grow gradually from $300\,$K to $60\,$K and then, quite unexpectedly, splits into a narrower Drude peak below about $100\,$cm$^{-1}$ and a broad peak at higher frequencies with a dip in between where the optical conductivity is suppressed. This splitting of the low-frequency peak resembles those from the spin density wave gap in pnictides and the hybridization gap in heavy fermion materials and is accompanied with the ferro-helimagnetic transition in MnP. While the above unpolarized data reveal interesting features at low temperatures, we should note that the analysis may not be accurate at low temperatures due to anisotropy. To avoid possible misinterpretation due to optical anisotropy and reveal the true origin of the gap feature, it is necessary to perform polarized measurements.\\

\begin{figure}[h]
\centering
\includegraphics[width=0.65\textwidth]{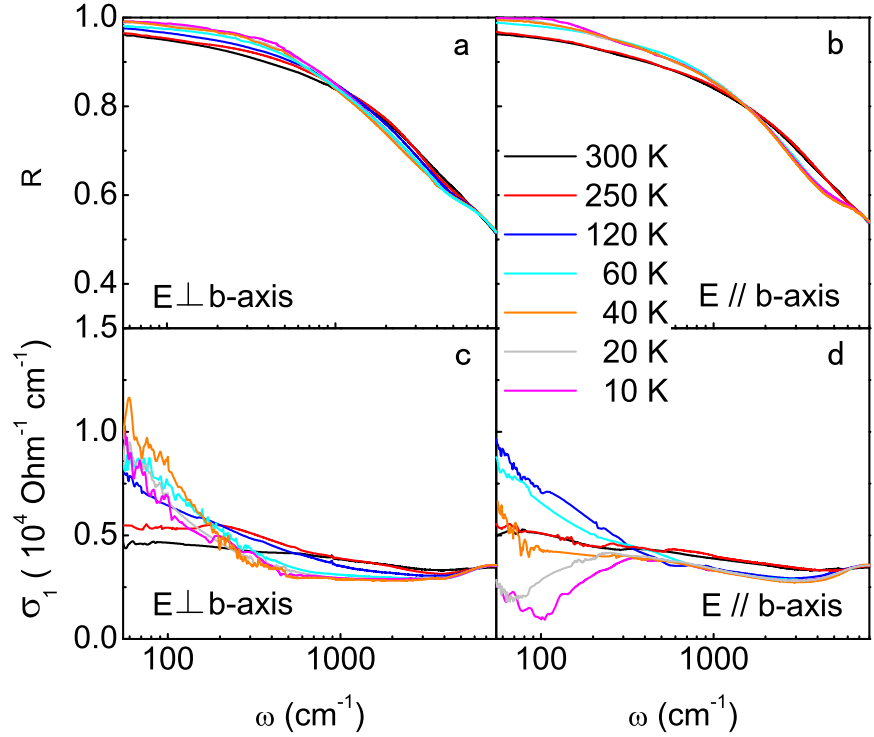}
\caption{\label{fig3}(Color online) \textbf{Polarized optical reflectivity and conductivity of MnP.} (a,b) Temperature dependence of the reflectivity spectra $R(\omega)$ for $\textbf{E}\perp b$ and $\textbf{E}\parallel b$; (c,d) Temperature dependence of the real part of the optical conductivity $\sigma_1(\omega)$ for $\textbf{E}\perp b$ and $\textbf{E}\parallel b$.}
\end{figure}

\begin{figure}[h]
\centering
\includegraphics[width=0.6\textwidth]{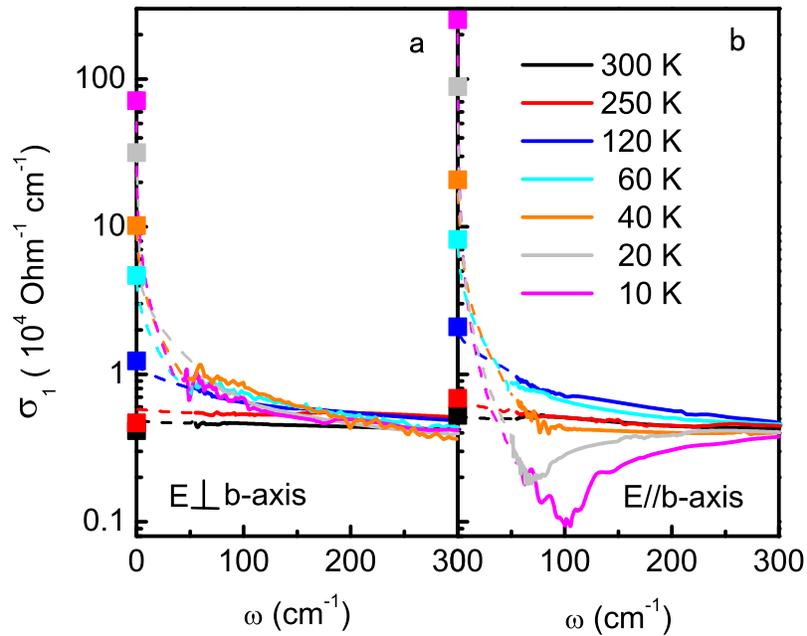}
\caption{\label{fig4}(Color online) \textbf{Low-frequency optical spectra in an enlarged scale for (a) $\textbf{E}\perp b$ and (b) $\textbf{E}\parallel b$}. The dashed lines represent the extrapolation below $55\,$cm$^{-1}$ using the Hagen-Rubens relation and the Kramers-Kronig transformation. The solid squares at zero frequency indicate the corresponding $dc$ conductivity $\sigma_{dc}=1/\rho_{dc}$ from the resistivity measurements.}
\end{figure}

\textbf{Polarized optical spectra.}
Figure \ref{fig3} shows the polarized reflectivity and optical spectra from $55\,$cm$^{-1}$ (limited by the reduced intensity of polarized light) to $8000\,$cm$^{-1}$ for $\textbf{E}\perp b$ and $\textbf{E}\parallel b$, respectively. The spectra in both directions merge together above $6000\,$cm$^{-1}$ and are the same as the unpolarized data for higher frequencies. Strong anisotropy only appears below $6000\,$cm$^{-1}$ and in the helimagnetic phase at low temperatures. Inside the helimagnetic phase, the gap opening is clearly seen along the $b$-axis but not visible in the perpendicular direction. Figure \ref{fig4} shows the enlarged plot for the low-frequency spectra. The dashed lines represent the low-frequency extrapolation using the Hagen-Rubens relation and the Kramers-Kronig transformation. We see that the extrapolated low-frequency optical conductivities agree well with the measured $dc$ conductivity at all temperatures, confirming the validity of our analysis for the polarized data.

To explore the properties of the charge carriers, we analyze the polarized $\sigma_1(\omega)$ using a combined multi-component Drude and Lorentz formula,
\begin{equation}
\nonumber\sigma_{1}(\omega)=\sum_i\frac{\omega_{p,i}^{2}}{4\pi}\frac{\gamma_{i}}{\omega^{2}+\gamma_{i}^{2}}
+\sum_j\frac{\Omega_{j}^{2}}{4\pi}\frac{\omega^{2}\Gamma_{j}}{(\omega^{2}-\omega_{0,j}^{2})^{2}+\Gamma_{j}^{2}\omega^{2}},
\end{equation}
where the first term gives the Drude contribution from itinerant charge carriers with the plasma frequencies $\omega_{p,i}$ and the scattering rates $\gamma_{i}$, and the second Lorentz term originates typically from inter-band transitions located at $\omega_{0,j}$ (corresponding to the direct gap) with the strength $\Omega_j$ and the damping coefficient $\Gamma_j$. The above formula can be applied because MnP exhibits weak electronic correlations with a small $\gamma_{e}\approx 8.3\,$mJ/mol K$^2$ at ambient pressure. An extended Drude formula may be needed near the critical pressure, where exotic non-Fermi liquid behavior has been observed \cite{Cheng}.

\begin{figure}[h]
\centering
\includegraphics[width=0.55\textwidth]{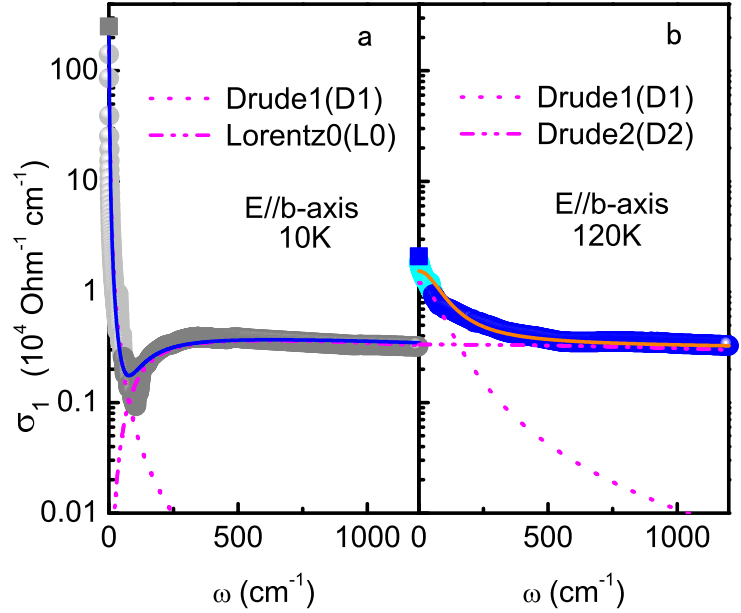}
\caption{\label{fig5}(Color online) \textbf{A tentative multi-component fit to the $b$-axis optical spectra at (a) $10\,$K and (b) $120\,$K.} The spectra show clearly two-component behavior. The overall fit (solid lines) is in rough agreement with the $dc$ conductivity and the extrapolated results (points) below $55\,$cm$^{-1}$ at both temperatures.}
\end{figure}

We first focus on the $b$-axis polarized data, which show interesting gap feature in the helimagnetic phase. Our analysis suggests that at least two Drude and two Lorentz (L1 and L2) components are needed in order to fit the experimental data above 60 K. The two Lorentz components describe the peaks centered at $\omega_{0,1}\approx 8100\,$cm$^{-1}$ and $\omega_{0,2}\approx 16000\,$cm$^{-1}$. In addition, two Drude components (denoted as D1 and D2) are required to fit the low and intermediate frequency regions in the ferromagnetic phase. Below 60 K, one of the components (D2) becomes gapped and turns into a Lorentz peak (L0), while the other gives rise to the narrow Drude peak (D1) below $100\,$cm$^{-1}$. The spectra are then composed of one Drude component (D1) and one Lorentz component (L0) in the helimagnetic phase. This indicates that there exist at least two types of charge carriers that behave very differently across the ferro-helimagnetic phase transition along the $b$-axis in MnP. To see this more clearly, we provide in Fig. \ref{fig5} a typical fit to the optical data above $55\,$cm$^{-1}$ that takes into consideration the measured $dc$ conductivity at zero frequency. While the results are consistent with the extrapolated spectra, the quantitative frequency-dependence below $55\,$cm$^{-1}$ may not be trusted at low temperatures ($T=10\,$K) because of the large difference between $\sigma_{dc}$ and $\sigma_1(\omega=55\,$cm$^{-1})$. Nevertheless, the multi-component model seems to capture at least the qualitative feature of the optical spectra in the whole temperature-frequency range.

For the polarized data with $\textbf{E}\perp b$, however, the two low-frequency terms are strongly overlapped. While we may expect some changes in the D2 component in the helimagnetic phase, it is impossible to be disentangled from the whole spectra because its contribution to the spectra at low frequencies ($\propto \omega_{p,2}^2/\gamma_2$) is two or three orders of magnitude smaller than that of the D1 component. As a matter of fact, we find that the low-frequency spectra can be equally well fitted with either two Drude terms or a Drude term plus a Lorentz term without affecting the fit at higher frequencies. This issue is avoided in the $b$-axis data due to the rapidly narrowed width of the D1 Drude peak in the helimagnetic phase, which reveals the change in the D2 component at around $100\,$cm$^{-1}$. This direction-dependence in the D1-component is due to the special ferromagnetic alignment of the Mn-spins along the $b$-axis in the helimagnetic phase at ambient pressure. We hence focus on the $b$-axis data in the following discussions.

\begin{figure}[h]
\centering
\includegraphics[width=0.65\textwidth]{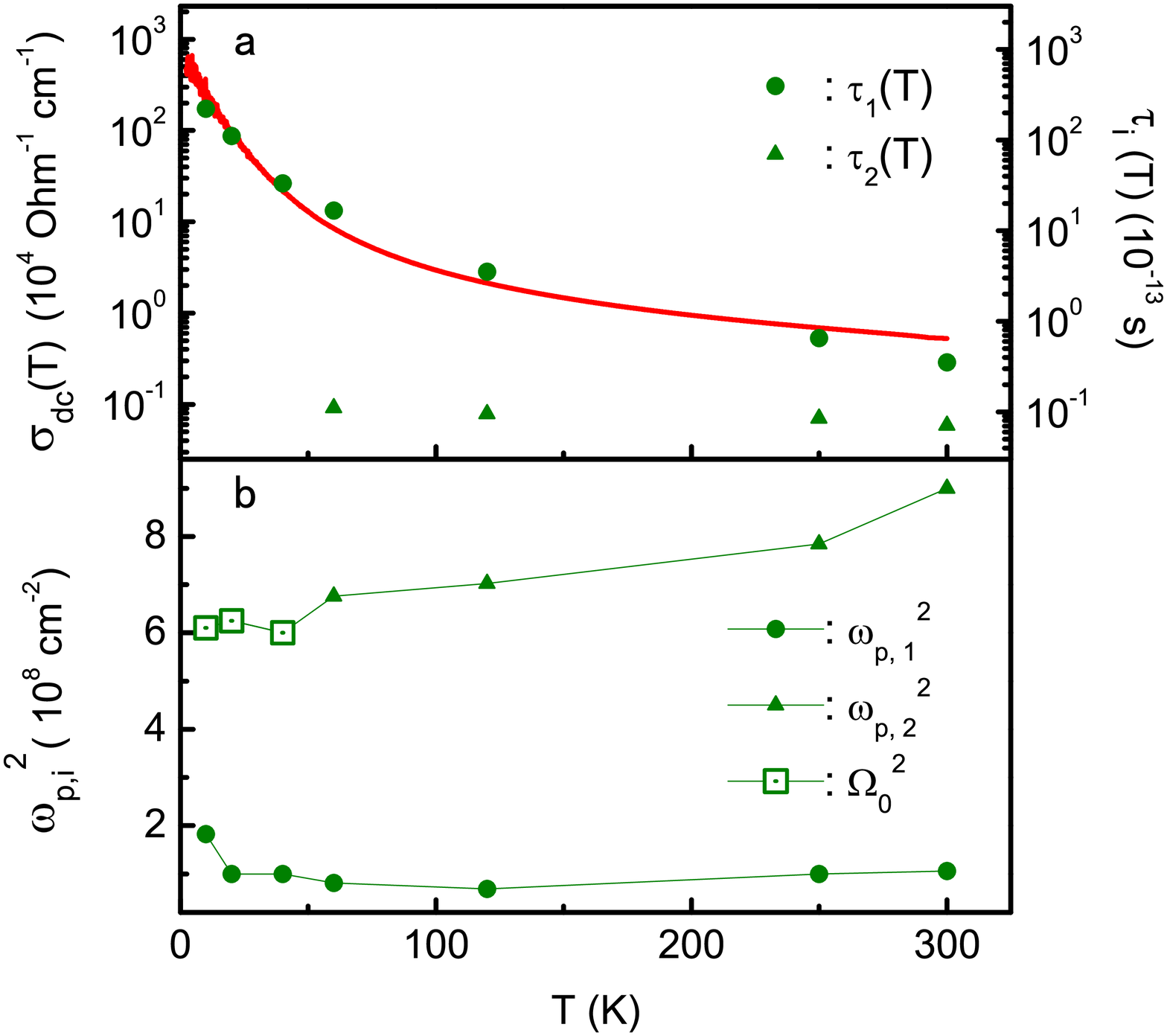}
\caption{\label{fig6}(Color online) \textbf{Fitting parameters for the two low-frequency components in the $b$-axis optical spectra.} (a) Temperature dependence of $\tau_i$ and its comparison with the $b$-axis $dc$ conductivity (solid line). (b) Temperature dependence of the corresponding $\omega_{p,i}^{2}$ in the ferromagnetic phase and its comparison with $\Omega_0^2$ in the helimagnetic phase.}
\end{figure}

To gain further insight into the properties of the charge carriers, we compare in Fig. \ref{fig6} the derived parameters as a function of temperature for the two low-frequency components based on the $b$-axis data. For completion, we also show the fitting parameters at low temperatures, even though their exact values could be influenced by the lake of experimental data below $55\,$cm$^{-1}$. We see that the two types of carriers have very different lifetimes calculated using $\tau_i=1/\gamma_i$. One has a long life time ($\tau_{1}$) that increases from $3.5\times10^{-14}\,$s to $2.2\times10^{-11}\,$s as temperature is lowered from 300$\,$K to 10$\,$K, and the other has a much shorter lifetime ($\tau_{2}$) of the order of $9\times10^{-15}\,$s that remains almost temperature independent. A simple comparison with other materials suggests that the former ($\tau_1$) behaves like a good metal ($\tau>10^{-13}\,$s) at low temperatures, while the latter ($\tau_2$) is strongly damped resembling those in other quasi-1D 3d/4d metals, e.g., TaS$_3$ ($\tau\approx 1.7\times10^{-15}\,$s) and NbSe$_3$ ($\tau\approx 0.9\times10^{-15}\,$s) \cite{Nakahara}. The small $\tau_2$ reflects the effect of electronic or magnetic correlations which become more significant at lower dimensionality. As already shown in Fig. \ref{fig5}b, the low-frequency conductivity is dominated by the long-lifetime carriers (D1) already at 120 K. This is further supported by the overall agreement between the $dc$ conductivity and $\tau_1$ (green solid circles) in Fig. \ref{fig6}a. The relative contribution of the short-lifetime carriers (D2) to the conductivity gets significantly less important as $\tau_2/\tau_1$ decreases rapidly with lowering temperature even before they are gapped in the helimagnetic phase. Figure \ref{fig6}b compares the plasma frequencies of both carriers. The strength $\Omega_0^2$ of the low-frequency Lorentz component (L0) in the helimagnetic phase is close to the value of $\omega_{p,2}^2$ of the second Drude term (D2) at $60\,$K. This indicates that they originate from the same electrons, namely the short-lifetime carriers that become gapped in the helimagnetic phase, turning the broad Drude peak into a Lorentz form in the optical spectra. We therefore conclude that there exist two distinct types of charge carriers in MnP which behave very differently and may be responsible for the transport (long-lifetime carriers, D1) and magnetic  (short-lifetime carriers, D2) properties, respectively.

\textbf{First-principles calculations.}
To understand the microscopic origin of both carriers, we performed the first-principles calculations for MnP using the full-potential augmented plane-wave and local orbital methods, as implemented in the WIEN2k code \cite{Blaha} and the non-collinear WIENNCM code \cite{Laskowski}. The Perdew-Burke-Ernzerhof generalized gradient approximation (GGA) was used for the exchange-correlation functional. The material has the space group $Pnma$ with an orthorhombic crystal structure and the lattice parameters are taken as $a=5.26\,${\AA}, $b=3.17\,${\AA} and $c=5.92\,${\AA} from previous work \cite{Gercsi}. For the ferromagnetic state, we obtain similar results with both codes, providing a consistency check for our numerical calculations. The helimagnetic state was calculated with the propagation vector $Q_h=(0,0,0.117)$ as observed in experiment \cite{Felcher}. The derived specific heat is about $4.4\,$mJ/mol K$^2$, close to the experimental value of $\gamma_{e}\approx8.3\,$mJ/mol K$^2$, indicating weak electronic correlations  and providing a further justification of our GGA calculations.

\begin{figure}[h]
\centering
\includegraphics[width=0.65\textwidth]{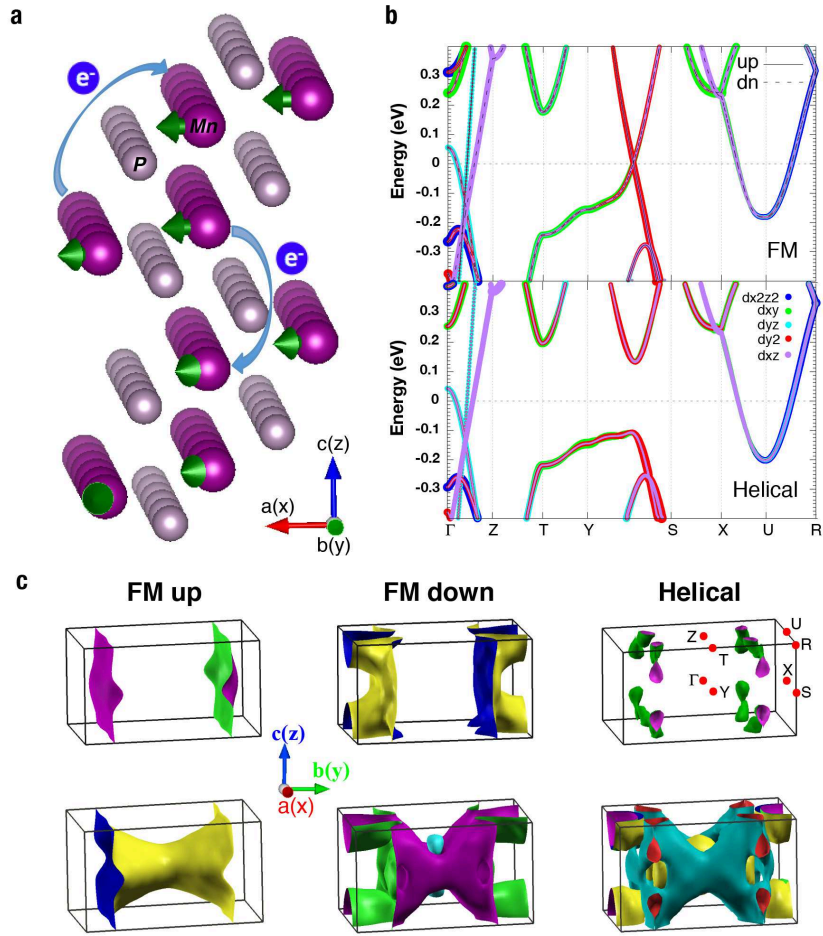}
\caption{\label{fig7}(Color online) \textbf{Band structures and Fermi surface topology of MnP.} (a) Illustration of the crystal structure and the helimagnetic order. The green arrows point to the directions of the ordered moments; the light blue arrows with the symbol "e$^-$" denote the electron hopping between Mn-ions, implying the itinerant character of the Mn $d$-electrons. (b) Comparison of the band structures in the ferromagnetic and helimagnetic states. (c) Comparison of the Fermi surfaces for the ferromagnetic and helimagnetic states. The top panels show the quasi-1D Fermi sheets (mainly of $d_{y^2}$-character) in the ferromagnetic state and their suppression in the helimagnetic state. The bottom panels show the rest parts of the Fermi surfaces from other orbitals.
}
\end{figure}

Figure \ref{fig7} shows the resulting band structures and Fermi surfaces for both the ferromagnetic and helimagnetic states in the unfolded Brillouin zone. In the ferromagnetic phase, the major Fermi surfaces have an interesting topology and are composed of one flat sheet and one sheet of the cylindric shape for each spin component. Both flat sheets show weak dispersion within the $xz$ plane and originate from the two intersecting bands along the Y-S direction in the Brillouin zone as shown in Fig. \ref{fig7}b. They are formed mainly by the Mn $d_{y^2}$-orbital that extends along the one-dimensional Mn chain, namely the $b$-axis, as illustrated in Fig. \ref{fig7}a. The two cylindric Fermi surfaces contain other Mn $d$-orbitals and exhibit anisotropic 3D character. The band structures of the helimagnetic phase show little changes except along the Y-S direction, where the two intersecting bands hybridize and open a small gap, owing to the helimagnetism that causes a mixture of the two spin components in neighboring Mn chains. In the optical measurement, this suppresses part of the optical spectra and gives rise to the dip feature as seen in Fig. \ref{fig4}b, in resemblance of those from the spin density wave gap in pnictides and the hybridization gap in heavy fermion materials. As a consequence, the Fermi surfaces show dramatic changes in the helimagnetic phase. The two flat sheets disappear completely, leaving a number of small pockets spreading at the corners of the Brillouin zone. These small pockets would be suppressed if the Coulomb interaction is taken into account. The cylindrical Fermi surfaces are also modified and become more 3D like. Our results are in good correspondence with the optical analysis and indicate that the short-lifetime carriers may originate mainly from the Mn 3$d_{y^2}$-orbitals and have flat Fermi surfaces, while the long-lifetime carriers contain the contributions from other Mn $d$-orbitals.

The existence of two types of charge carriers with distinct properties results naturally from the particular crystal structure of MnP and is therefore robust against pressure or other external tuning parameters without altering the crystal structure \cite{Takase3}. The fact that the two types of carriers dominate separately the transport and magnetic properties provides a possible simplification for our understanding of its exotic and complicated physics \cite{Cheng,Takase1}. For example, the helimagnetic order may be determined primarily by the Mn 3$d_{y^2}$-orbitals and the Fermi liquid behavior in the helimagnetic phase at ambient pressure may be largely attributed to the long-lifetime carriers from other orbitals. At the helimagnetic transition, the small kink in the resistivity perpendicular to the $b$-axis could be understood as the magnetic scattering of the long-lifetime carriers by the helimagnetic fluctuations in the $xz$-plane. Our results also put a strict constraint on any serious theoretical investigations. A proper microscopic model should contain at least two bands of different characters and treat correctly the quasi-1D nature of the Mn 3$d_{y^2}$-band at low energies. Taking this as a starting point, we speculate that at higher pressures, the superconducting and helimagnetic properties of MnP may also be attributed to the interplay between the two types of carriers, causing non-Fermi liquid behavior and pairing condensation near the helimagnetic quantum critical point, in resemblance of that in the heavy fermion superconductors.

\section*{Discussion and conclusions}
The discovery of superconductivity in MnP has stimulated many experimental and theoretical studies in past few years. Despite of these and earlier efforts, some fundamental issues remain unclear. Our work made contributions in the following aspects:
\begin{itemize}
\item Our work provides the first optical data that may help to reveal the peculiar charge dynamics of MnP. The optical measurements were very challenging due to the small size of the superconducting single crystals. So far, no other probes has been able to provide such information and ARPES has failed to produce useful data, possibly due to the lack of a good cleaved surface.

\item The temperature dependence of the optical data reveals the coexistence of two types of charge carriers with very different lifetimes. This can be clearly seen from the spectra alone and is therefore model-independent. The model analysis only helps to provide a better illustration of the two components and a quantitative estimate of the magnitude of their lifetimes. 

\item Crossing the helimagnetic transition, the Fermi surfaces of the short lifetime carriers become gapped, which coincides with the gap opening of the $d_{y^2}$-band in the GGA band structure calculations. While one should not expect the simple GGA to give a quantitative solution of such a correlated material, the connection provides an insight on the origin of the short lifetime carriers and the helimagnetic instability. As a matter of fact, it has been shown that the special band crossing along the Y-S direction as in Fig.~\ref{fig7}b is particularly susceptible to non-collinear instabilities \cite{Lizarrage2004}. Helimagnetism in MnP has often been modeled based on a fully localized picture \cite{Huber, Yamazaki, Zieba}. However, combining our experimental and numerical results suggests that the helimagnetic order in MnP might have an itinerant origin, in contrast to the theories based on fully localized spins. Our work therefore provides useful insight on an important and fundamental question in MnP.

\item Based on the above analysis, one may expect that a realistic model of MnP must take into account the two-component feature of the charge carriers. While the short lifetime carriers are primarily responsible for the helimagnetic instability, superconductivity can only emerge out of the long lifetime carriers, possibly with pairing glues coming from the helimagnetic spin fluctuations.
\end{itemize}

To summarize, we report optical investigations on the single crystal MnP and find that the optical conductivity can only be well understood with a multi-component Drude and Lorentz formula. In combination with band structure calculations, our analysis reveals the existence of two types of charge carriers with distinctly different characters. The short-lifetime charge carriers become gapped along the $b$-axis in the helimagnetic phase and are primarily responsible for the magnetic properties, while the long-lifetime charge carriers are more 3D-like and provide major contributions to the transport properties such as the dc conductivity. This puts a microscopic constraint for understanding the basic physics of MnP.

\section*{Methods}
\textbf{Sample preparation and measurements.} Four needle-shaped single crystals, three with an average size of about $2\times0.3\times0.08\,$mm$^{3}$ and one about $2.1\times1.4\times0.5\,$mm$^{3}$, have been grown using selenium-flux method and used in this experiment \cite{Cheng}. The $dc$ resistivity  $\rho_{dc}$ and the specific heat $C_p$ were measured with the four-probe method and the thermal relaxation method, respectively, in a Quantum Design physical property measurement system (PPMS). The optical measurements were performed on the Fourier transform spectrometers (Bruker 113 v and 80 v/s) using an $in\ situ$ gold and aluminium overcoating technique.


\section*{Acknowledgements}
This work is supported by the National Science Foundation of China (Nos. 11574358, 11522435, 11774401), the 973 project of the Ministry of Science and Technology of China (Nos. 2015CB921300, 2015CB921303), the National Key Research and Development of China (Nos. 2016YFA0300303, 2017YFA0303103), and the Strategic Priority Research Program of the Chinese Academy of Sciences (Grant No. XDB07020200).\\

\section*{Author contributions statement}
P.Z., Y.Y., and J.L.Luo supervised the project; P.Z., W.W., G.X., J.L.Lv, P.W., and F.K.L. performed the measurements; Y.J.X. and Y.Y. did the theoretical calculations; all authors discussed the results; P.Z. and Y.Y. wrote the paper.\\

\section*{Additional information}

\textbf{Competing financial interests:} The authors declare no competing financial interests.

\end{document}